# A universal deep learning strategy for designing high-quality-factor photonic resonances

Xuezhi Ma[1] †, Yuan Ma[1]†, Preston Cunha[1], Qiushi Liu[2], Kaushik Kudtarkar[1], Da Xu[2], Jiafei Wang[1], Ming Liu[2], M. Cynthia Hipwell[1,] and Shoufeng Lan[1*]

1 Department of Mechanical Engineering, Texas A&M University, College Station, TX 77840, USA.

2 Department of Electrical and Computer Engineering, University of California, Riverside, CA 92521, USA.

† These authors contributed equally.

Author e-mail address: shoufeng@tamu.edu

**Abstract**

Resonance is instrumental in modern optics and photonics for novel phenomena such as cavity quantum electrodynamics and electric-field-induced transparency. While one can use numerical simulations to sweep geometric and material parameters of optical structures, these simulations usually require considerably long calculation time (spanning from several hours to several weeks) and substantial computational resources. Such requirements significantly limit their applicability in understanding and inverse designing structures with desired resonance performances. Recently, the introduction of artificial intelligence allows for faster predictions of resonance with less demanding computational requirements. However, current end-to-end deep learning approaches generally fail to predict resonances with high quality-factors (Q-factor). Here, we introduce a universal deep learning strategy that can predict ultra-high Q-factor resonances by decomposing spectra with an adaptive data acquisition (ADA) method while incorporating resonance information. We exploit bound states in the continuum (BICs) with an infinite Q-factor to testify this resonance-informed deep learning (RIDL) strategy. The trained RIDL strategy achieves high-accuracy prediction of reflection spectra and photonic band structures while using a considerably small training dataset. We further develop an inverse design algorithm based on the RIDL strategy for a symmetry-protected BIC on a suspended silicon nitride photonic crystal (PhC) slab. The predicted and measured angle-resolved band structures show minimum differences. We expect the RIDL strategy to apply to many other physical phenomena which exhibit Gaussian, Lorentzian, and Fano resonances.

**Introduction**

As one of the most fundamental building blocks of the physical world, resonance is ubiquitous, ranging from daily phenomena such as musical instruments to cutting-edge science and technologies like the Fabry-

Perot resonator in quantum systems [1]. Resonances exhibit enhanced amplitudes of energy modulation under a periodic load at frequencies near their natural frequencies, storing large energy density. Specifically, in the field of optics, optical resonators such as optical ring resonator [2], Mie resonator [3], photonic crystal cavities [4], etc., allow a beam of light to circulate and harness photonic energy in the form of resonances. The ability of a resonator to store energy can be represented by the ratio of the initial energy stored in the resonator to the energy lost in one radian of the cycle of oscillation, i.e., the quality factor (Q-factor).

Thanks to its utmost importance in science and engineering fields, resonances have been studied by generations of scholars. Galileo Galilei developed the Lorentzian formula for the resonance effects in musical strings in 1602. Three hundred years later, Ugo Fano reported a more universal resonance model in 1961, now known as Fano resonances [5]. In contrast to Lorentzian resonances, the Fano resonances exhibit asymmetry with respect to energy, which comes from the coupling between a discrete localized state and a background continuum state. Fano resonances can degenerate into Lorentzian resonances when its asymmetry parameter (also known as the Fano parameter) approaches infinity [6]. Fano resonances have been found in plasmonic nanoparticles [7], photonic crystals [8], metasurfaces and metamaterials [9, 10], optical cavities [11], gratings [12], etc., and have been applied in sensors [13], nano-lasing [14], modulation [15], and light buffering and storage [16]. Recently, an infinite Q-factor Fano resonance has been observed on BICs with localized bounded states coexisting in the same energy with unbounded continuum states [17]. With their infinite Q-factor, optical BICs can harness a large amount of energy and enhance light-matter interactions significantly [8].

The resonances of optical resonators are very sensitive to their geometry and material parameters. Understanding their relationship is crucial for the design of resonators, especially high Q-factor resonators. Numerical simulations including the finite element methods (FEM) and finite-difference time-domain (FDTD) methods are capable of calculating resonances with given sets of parameters. These simulations, however, usually require a relatively large amount of time (spanning from several hours to several weeks) and substantial computational resources, thereby limiting their capacity to provide a thorough understanding of the underlying relationship between resonance parameters and design targets. For critical simulation tasks like inverse design, such methods become insufficient [18]. Recently, artificial intelligence (AI) algorithms including machine learning and deep learning have been introduced to the field of resonator design and have significantly accelerated the design progress for multiple purposes. With pre-acquired training data, a neural network can be trained with a backward propagation algorithm to predict an optical response typically within less than 1s [18-21]. This great reduction in computational time using deep learning algorithms enables large-scale computation and inverse designs. Deep learning algorithms have been applied to predict the reflection or transmission spectra of many different metasurfaces or structures, such as plasmonic core-shell particles [22], plasmonic chiral metamaterials [23], gratings [24], metalens [25], photonic crystals [26], etc. In these applications, geometries and properties of the optical structures were treated as inputs of a neural network and the reflection or transmission spectra with respect to energy as outputs. These state-of-the-art approaches, however, are still unable to predict reflection or transmission spectra with high-Q resonances (ultra-sharp peaks on spectra) [27]. The capacity of predicting such spectra with ultra-sharp peaks is, in fact, one of the most crucial evaluation criteria for resonator designs. Without accurate knowledge of the resonance parameters, AI-based resonator inverse design is impossible to achieve. Some valuable attempts including interpolation have been reported [28]. These approaches, unfortunately, are still unable to provide accurate predictions [29].

**Results and discussion**

In this article, we report a RIDL strategy for accurate ultra-high Q-factor resonance prediction and inverse design. Inspired by physics-informed machine learning [30, 31], the RIDL strategy treats Fano resonances as the most fundamental component for generating resonance spectra. Unlike the traditional end-to-end deep learning neural networks that establish the direct mapping between the inputs and outputs, the RIDL strategy explicitly imposes prior-exist laws of resonance phenomenon into the neural network as regularization agents. In other words, excluding some impossible solutions for the backward propagation algorithm can reduce the complexity, greatly reducing the size of the training dataset needed to provide accurate connections between inputs and outputs. **Figure 1a** shows a general structure of this RIDL strategy. For the forward direction, resonances that exhibit Fano, Lorentzian, or Gaussian peaks are treated as the most fundamental components and imposed into the deep learning neural network. With the prior-acquired information, such as the equation of a Fano peak, a resonance can be decomposed from the whole spectrum by a decomposer and transformed into a set of analytical parameters. The decomposed resonances are classified into categories with the assistance of resonance field information (electric-fields, magnetic-fields, etc.) to separately train regression neural networks for resonance parameters predictions. Finally, resonances are regenerated from predicted resonance parameters and combined into predicted spectra with resonances.

We chose a suspended silicon nitride ($Si_3N_4$, refraction index 2.02) photonic crystal (PhC) slab with $C_4$ symmetry as the optical resonance platform (**Figure 1b**) to benchmark the RIDL strategy. The oblique incident broadband light can be trapped and resonate in the photonic crystal slab, generating a reflection spectrum with a high-Q Fano lineshape. Further, the symmetry-protected BIC on the Gamma point ($\Gamma$) in the band structure (on-$\Gamma$ BIC) is a very special phenomenon supported by this PhC slab [17]. The on-$\Gamma$ BIC can be excited by normal incident light with vanished Fano lineshape in its reflection spectrum (indicative of an infinite Q-factor), which means all the energy is trapped in the BIC mode without escaping. The vanished Fano can be found by sweeping geometric parameters or incident angles in the band structure of the PhC to find the vanishing Fano lineshape by observing the trend of a series of its adjacent Fano lineshapes. To design a BIC, the first task is building the relationship between the inputs (including geometric parameters, incident angle, and wavelength of light) and its reflection spectra. **Figure 1c** shows the reflection spectrum of a suspended silicon nitride PhC slab with a pitch of a = 500 nm, a radius of periodic holes of r = 132 nm, a slab thickness of h = 175 nm, and under an oblique incident angle of β = 1.1° with P-polarization (transverse magnetic modes). **Figure 1d** shows the predicted spectrum by RIDL, which exhibits high accuracy when compared with the ground truth (**Figure 1c**). It should be noted that the Q-factors of the resonances in Figure 1c and 1d are as high as 69 (TE), 1353 (TM1), 132 (TM2) and 6177 (TM3). RIDL was also compared with an end-to-end machine learning algorithm where a neural network was trained with geometric parameters as inputs and reflection coefficients at different energies as outputs (**Figure 1e**) using the exact same training dataset. Training data was obtained using the adaptive data acquisition (ADA) method, which can sufficiently describe spectra with resonances using minimum data points and significantly reduce computational time in acquiring spectra with high resolution (more information about ADA is included in Methods). While training the end-to-end neural network, each training spectrum was comprised of simulated data with a very fine energy resolution ($\Delta E < 8 \times 10^{-7}$ eV, capable of describing Q-factor as high as $10^6$) for accurate Fano resonance descriptions, resulting in an overall training dataset size of more than 1 million points. Even with such a spectrum resolution, the resultant spectrum was still less than ideal. The end-to-end machine learning model can predict the TE1 resonance which has a relatively broader lineshape (lower Q-factor) better than the other resonances with sharp peaks. There are several explanations for the unsatisfying performance of the end-to-end approach: (a) The training dataset is only about 4,000 cases, about 10% of a typical machine learning approach for photonic crystals (Comparisons see Supplementary Figure 1). (b) In the end-to-end approach, data points on the resonances with high Q-factors are often out-weighted by data points elsewhere. (c) Fano resonances with different Q-factors require different orders of complexity in the network itself: sharper peaks will require higher orders of non-linearity in the network. A nonlinear regression network with adaptive orders of non-linearity for varying Q-factor resonances is required.

**Resonance-informed deep learning (RIDL) strategy.** Rather than relying on the end-to-end approach to establish a direct connection between the inputs and reflection spectra, the RIDL strategy decomposes each Fano peak from the reflection spectra and describes the Fano equations accurately with the Fano equation:

$$\sigma(E; E_0, \Gamma, \delta) = D^2 \frac{(q+\Omega)}{(1+\Omega^2)} \qquad (1)$$

where E is the energy of light, q = cot $\delta$ is the Fano parameter, $\delta$ is the phase shift, $\Omega = 2(E - E_0)/\Gamma$, where $\Gamma$ and $E_0$ are the resonance width and resonance energy respectively, and $D^2 = 4 \sin^2 \delta$. The Fano equation allows RIDL to directly recognize resonance peaks of different Q-factors without high order complexity. In comparison to the end-to-end approach, the RIDL strategy compressed redundant degrees of freedom of the training dataset and allowed for an efficient and integrated description of the Fano lineshape. The Fano resonance information, serving as the latent information in the RIDL strategy, helped regularize the neural networks' parameter space, allowing for much faster computational times and only requiring a very small training dataset.

The spectrum shown in **Figure 2a** is a simulated reflection response from a $Si_3N_4$ PhC slab determined using the ADA method (see Methods). The $Si_3N_4$ is a dielectric material in the wavelength range of interest (600 nm to 900 nm) and the refractive index can be treated as a constant without a loss (image part of the refractive index). In the spectrum, the P-polarized incident light excited three transverse magnetic (TM) bands and one transverse electric (TE) band in the wavelength range of interest for the five lowest TM or TE bands supported by the photonic crystal. We choose these four bands as the bands of interest (color-coded as green, red, pink, and yellow for TE1, TM1, TM2, and TM3) whereas other higher-order bands were classified into the "others" category (color-coded as gray) (**Figure 2c**).

The decomposition algorithm first separated the Fano peaks (**Figure 2c**) from the background spectrum (**Figure 2b**). The background spectrum is a smooth spectrum without higher-order nonlinearity and is suitable for end-to-end machine learning regression (**Figure 2f**). In the RIDL network, the radius of holes (r), slab thickness (h), oblique incident angle (β), and wavelength of each point in a spectrum are treated as inputs. Pitch size (a = 500 nm) is fixed in this work to downsize the number of parameters to reduce the complexity. For cases with different pitch sizes, the scaling strategy can be applied to varying the parameter of pitch size [32]. The extracted Fano peaks are assembled in a Fano pool that contains all Fano peaks and their corresponding electric field (E-field) distributions from ~4,000 training datasets with uniformly distributed random parameters.

The electric fields in the normal direction ($E_z$) at the top surface of the PhC slab unit cell are simulated using COMSOL at each Fano peak. **Figure 2h** shows the semi-supervised convolution neural network (CNN) to classify each Fano from the pool (~30,000 Fano peaks) into five categories (TE1, TM1, TM2, TM3, or others). We manually labeled ~700 spectra to identify 600 Fano resonance peaks of each category (3,000 peaks in total) from the entire Fano pool. 200 Fano peaks of each category (1,000 in total) were randomly selected to train a multi-inputs CNN classification network which takes r, h, β, $E_0$, $\Gamma$, and $\delta$, in addition to E-field distributions, as inputs and the five categories as outputs (**Figure 2h**). The 41-pixel by 41-pixel mapping of E-field distributions before the peak (E = $E_0 - \Gamma/8$), on-peak (E = $E_0$), and after the peak ($E = E_0 + \Gamma/8$) were stored in 3 channels of a figure to serve as the E-field distribution input of the CNN. The CNN consists of two sets of convolution layers, batch normalization layers and max-pooling layers, followed by a concatenation layer that combines the convolution branch and parameter input branch. The trained CNN classification network was validated with 100 Fano peaks from each category (500 in total) and tested by 300 Fano peaks for each category (1,500 in total), yielding a 96% validation accuracy and 95% test accuracy after 5,800 iterations (more details see Supplementary Information). The parameters of the four different Fano peaks were then used to train four parallel regression neural networks where r, h, and β are treated as the inputs and $E_0$, $\Gamma$, and δ as the outputs. The predicted Fano parameters were used to reconstruct Fano peaks with high accuracy and low computation resource consumption. The background spectrum that is predicted by the background end-to-end regression network (**Figure 2f**) and Fano peaks from the parallel Fano regression networks (**Figure 2g**) were finally combined to construct the predicted reflection spectrum. The assembled spectrum (**Figure 2d**) showed a minimum difference from the ground truth (**Figure 2a**) when the fifth Fano peak (other Fano category) is ignored.

**Figure 3a** and **Figure 3b** show the mean-squared error (MSE) loss while training the regression networks for TE1, TM1, TM2, TM3, and the background spectra. It is clear that the training error and validation error decreased significantly after 60,000 and 1,000 epochs of Fano peak networks and background networks, respectively. **Figure 3c** shows the classification accuracy and confusion matrix for the Fano resonance classifier network. With only 200 labeled Fano peaks for each resonance category to train the classification network, a high accuracy (95%) was achieved on the test set with low bias. This trained network served as the classifier to label the rest of the decomposed Fano peaks which further served as the training data for the regression networks. The combination of the classifier and regression network enables the RIDL network to learn and predict the resonances in a semi-supervised fashion.

The classifier in the RIDL network ensures that each regression network is trained with the right training dataset without category mismatching: with different geometry parameters and oblique angle of the incident light, the relative position of the four targets Fano resonances with respect to energy changes. **Figure 3d** shows a case where the TM1 resonance is situated between the TE1 and TM2 resonances. Both TE1 and TM2 are relatively broad, and the two resonances interact with each other to form a saddle-shaped complex resonance on the reflection spectrum, whereas the TM1 peak is embedded on the saddle with a very sharp peak (higher Q-factor). This case is also one of the many cases that demonstrated the robustness of RIDL to be able to identify, separate, and predicted all target resonances (more comparison between prediction results from RIDL and simulation ground truth are included in supplementary information). It is also worth mentioning that the highest intensity of the TM1 peak in the ground truth did not reach 1 because of the limited resolution of the numerical simulation to describe such a high Q-factor (~7008) Fano resonance.

However, on the spectrum predicted by RIDL, the analytically parameterized description of the resonance allowed the spectrum to be represented at any required resolution.

**Inverse design of a BIC.** Once the RIDL is well-trained for forward prediction, this network can be used for the inverse design of photonic crystals with specific band structure requirements. As previously mentioned, a symmetry-protected BIC vanishes in the reflection spectrum with infinite Q-factor because no reflected light can escape. The best way to design a PhC with BIC is relying on its adjacent Fano peaks with increasing Q-factors in the angle-resolved reflection band structure. By sweeping the input incident angle in the trained RIDL, a series of reflection spectra of the same PhC can be predicted to form an angle-resolved reflection band structure. Different from the numerical simulations, the angle-resolved resolution of the reflection band structure generated by the RIDL has almost no resolution limit, thanks to the small amount of computation time in generating predictions.

The inverse design algorithm is developed based on a non-linear multivariate optimization method. To demonstrate the inverse design algorithm, the following constraints were imposed while inversely designing a PhC with BIC:

1) The radius of the PhC is larger than $r_{min} = 80$ nm. (For fabrication yields)
2) The thickness of the PhC slab ranges from $h_{min} = 200$ nm to $h_{max} = 300$ nm.
3) The first BIC is located at $\lambda^{TM1} = 700$ nm with a tolerance of 3nm.
4) The second BIC is located at $\lambda^{TM3} = 750$ nm with a tolerance of 3nm.

The objective function and constraint functions of the inverse design algorithm were defined as follows:

$$\begin{cases} \text{Minimize} \quad F(r,h,\beta_0) = C_1 \left( E_0^{TM1}(r,h,\beta_0) - \frac{hc}{\lambda^{TM1}} \right)^2 + C_2 \left( E_0^{TM3}(r,h,\beta_0) - \frac{hc}{\lambda^{TM3}} \right)^2 \\ \text{Subject to} \quad r > r_{min} \\ \qquad\qquad\quad h_{min} < h < h_{max} \end{cases} \quad (2)$$

where $C_1$ and $C_2$ are weights for the two BIC, h is Planck's constant, c is the speed of light in vacuum, and $\beta_0 = 0$ for on-$\Gamma$ BIC. This optimization problem is solved with an interior-point algorithm in MATLAB. Other optimization algorithms including genetic algorithms (GA) [33], gradient descent [34], and stochastic algorithms [35] can also be applied in specific inverse-design problems. It should be noted that since the RIDL strategy directly predicts the resonance parameters, this network is easier to incorporate in inverse design algorithms.

With relatively high tolerances, the inverse design algorithm can provide a design with integer numbers to satisfy the fabrication yields: r = 101 nm, and h = 238 nm. **Figure 4a** shows the band structure of the predicted design calculated by the MPB band solver [36]. TE1, TM1, TM2, and TM3 bands excited by P-polarized oblique incident light were marked with solid lines whereas the rest of the bands (dashed lines) can only be excited by S-polarized incident light and were not observed in the experiment [37]. **Figure 4b** shows the reflection band structure comparison generated by the numerical simulation (COMSOL) that serves as the ground truth (**left**) and by the RIDL strategy (**right**). It is worth noting that the resolution of the angle-resolved band structure generated by the RIDL is much higher than its numerical counterpart because of the big difference in computing speed.

A real device following this set of parameters was fabricated and the SEM image is shown in **Figure 4c.** The device is 100 pitches by 100 pitches (50 μm by 50 μm) of suspended PhC slab supported by an undercut silicon wafer (for more fabrication details see Methods). The band structure was measured via angle-resolved reflection spectroscopy (See Method and Supplementary). Two desired BIC modes (gradually vanished points) located at 700 nm and 750 nm can be clearly observed.

**Transfer learning.** While the current RIDL is developed with varying geometric parameters, other variabilities in design and even material of the PhC can be easily incorporated with transfer learning [38]. The transferability of RIDL is demonstrated here with the change of the refractive index of $Si_3N_4$. Depending on fabrication protocols, the refractive index of $Si_3N_4$ can vary between 2.0 and 2.4. The trained RIDL using 2.02 as the refractive index was transferred to predict PhC with a refractive index of 2.23. The transfer learning was achieved by floating only the last several layers of each of the regression networks on TE1, TM1, TM2, TM3, and the background spectra, and re-training these networks with only a very small set of training dataset that was numerically simulated with the refractive index of 2.23. Similarly, transfer learning of RIDL can allow the network to be trained to predict other geometric and material variations including pitch size, hole shape, material properties, etc. More information on transfer learning of RIDL is included in the Supplementary Information.

**Conclusion**

In light of the unsatisfactory prediction and inverse design of ultra-high Q-factor resonances using end-to-end deep learning approaches, we demonstrated here a Resonance-informed deep learning (RIDL) strategy. This strategy introduced the prior-exist knowledge of resonance equations as a regularization agent into the deep learning algorithm, allowing the deep learning neural network to better understand and predict resonances with ultra-high Q-factors. This RIDL strategy significantly reduced the required size of the training dataset, i.e., only approximately 10% of the typical amount of data is needed to guarantee high accuracy compared to traditional end-to-end deep learning algorithms. The predicted reflection spectra from RIDL accurately match their corresponding ground truths. An inverse design algorithm was further developed with the RIDL strategy to generate PhC slab designs that satisfy multiple constraints. A PhC slab with BIC modes at 700 nm and 750 nm was designed with this inverse design algorithm, validated by a device fabricated on $Si_3N_4$ following the inverse design result. The angle-resolved band structure of the fabricated PhC slab unambiguously follows both the ground truth and predicted band structure, satisfying all constraints imposed in the inverse design. Besides Fano resonances, the RIDL strategy also applies to other physical resonances such as Gaussian and Lorentzian resonances.

## Methods

### Sample fabrication

A $Si_3N_4$ layer of 300 nm thickness was deposited on commercial Si substrates by low-pressure chemical vapor deposition (LPCVD). The periodic PhC pattern with the pitch of 500 nm and hole radius of 101 nm was written on ZEP-520 photoresist and followed by the RIE-ICP cyclic dry etch strategy to transfer the pattern into the $Si_3N_4$ layer [39]. After removing the residual ZEP-520 photoresist by acetone and oxygen plasma cleaning procedure, the silicon (<100> oriented) beneath the PhC pattern was undercut by using KOH solution (30 wt. %) at 120°C to suspend the PhC slab. Finally, the modified RIE etching was applied without a mask to reduce the thickness of the slab to 238 nm.

### Optical measurement

Angle-resolved reflection spectroscopy was performed by a Fourier transform system to project the back focal plane of a 20X Mitutoyo Plan Apo Infinity Corrected Long WD Objective. A slit spectrometer (Princeton Instruments SP300) was placed at the K-plane for the band structure. The white light came from a halogen lamp and the illumination pattern in real space was cut by a pinhole to fit the device which has a limited size (see Supplementary Fig. S3).

### Adaptive data acquisition (ADA) method for high-resolution spectra

Since the spectra of the PhC with BIC have a wide range of peak widths, equal-spaced wavelength sweeping could not balance the need for precision and computational efficiency. Therefore, an adaptive data acquisition (ADA) method is developed in this paper to automatically refine the sampling mesh where a Fano resonance may exist. The refined criteria are based on interpolation error $e(E)$, which is defined as

$$e(E_i) = (R_0(E_i) - R_I(E_i))^2 \tag{3}$$

where $R_I(E_i)$ is the interpolated reflection coefficient with neighboring reflection coefficients of $E_i$ using piece-wise cubic interpolation. The detailed ADA structure is described here:

**Input:** Initial spectrum $R_{(0)}(E)$, where $E$ is uniformly sampled from $E_{min}$ to $E_{max}$. In this paper, 301 points (1nm/point) were sampled for each initial spectrum.

**Output:** Refined spectrum $R(E)$.

1   **Begin**
2       define an empty refine-rejection list
3       **for** t=1, 2, …, m, where m is the predefined number of refining iterations
4           calculate interpolation error $e(E_i)$ for all $R_{(t-1)}(E_i)$
5           find $E_i$ (count =n) that satisfies:
                (a). $e(E_i)$ are local maxima
                (b). $e(E_i) > threshold \times standard\_deviation(e(E_i))$
                (c). $E_i$ is not in the vicinity of any point on the refine-rejection list
6           **for** i=1, 2, …, n
7               calculate $R(E_i - l)$ and $R(E_i + r)$, where $l$ and $r$ are half of the current grid size on either side of $E_i$
8           assemble $R_{(t)}(E_i)$
9           evaluate each peak in $R_{(t)}(E_i)$, if the peak is prominent enough, its corresponding $E$ is recorded in the rejection list

With this ADA method, the numerical simulation achieved an immense reduction in computational time: the computational time of a spectrum with ADA is reduced by **~98.5%** when compared with a uniform sampling of $E$ with the same resolution for Fano peaks.

**Decomposition algorithm**

The spectrum numerically calculated with COMSOL was decomposed with the decomposition algorithm described here. While this paper adopts an optimization-based decomposition algorithm, other methods including 1D CNN and RNN could also potentially be applied.

The initial spectrum calculated from a numerical simulation can be expressed as

$$R = R(E) \tag{4}$$

Since this spectrum consists of a series of Fano resonances and a background reflection, R can be expressed as

$$R(E) = \sum_{i=1}^{n} \bar{\sigma}^{(i)}\left(E; E_0^{(i)}, \Gamma^{(i)}, \delta^{(i)}\right) + B(E) \tag{5}$$

where $\bar{\sigma}_i(E)$ is the $i^{th}$ normalized Fano resonance

$$\bar{\sigma}^{(i)}\left(E; E_0^{(i)}, \Gamma^{(i)}, \delta^{(i)}\right) = \frac{1}{4}\left[\sigma\left(E; E_0^{(i)}, \Gamma^{(i)}, \delta^{(i)}\right) - D^2\right] \tag{6}$$

The decomposition is performed one Fano resonance at a time. For the $i^{th}$ Fano resonance, the arc length of the residual spectrum is minimized with optimization algorithm:

$$\begin{cases} \text{Minimize} \quad r\left(E_0^{(i)}, \Gamma^{(i)}, \delta^{(i)}\right) = \int_{E_{min}}^{E_{max}} \sqrt{1 + \left(\frac{R^{(i)}(E) - \bar{\sigma}^{(i)}(E; E_0^{(i)}, \Gamma^{(i)}, \delta^{(i)})}{dE}\right)^2} \, dE \\ \text{Subject to} \quad E_{min} < E_0^{(i)} < E_{max} \\ \quad\quad\quad\quad\quad \Gamma^{(i)} > 0 \\ \quad\quad\quad\quad\quad 0 < \delta^{(i)} < \pi \end{cases} \tag{7}$$

where $R^{(i)}(E)$ is the current spectrum after $i - 1$ iterations of decomposition.

In the discrete form, the objective function can be written as

$$r\left(E_0^{(i)}, \Gamma^{(i)}, \delta^{(i)}\right) = \sum_{j=1}^{m-1} \sqrt{\left[\left(R^{(i)}(E_{j+1}) - \bar{\sigma}^{(i)}(E_{j+1}; E_0^{(i)}, \Gamma^{(i)}, \delta^{(i)})\right) - \left(R^{(i)}(E_j) - \bar{\sigma}^{(i)}(E_j; E_0^{(i)}, \Gamma^{(i)}, \delta^{(i)})\right)\right]^2 + \left(E_{j+1} - E_j\right)^2} \tag{8}$$

The reason this decomposition algorithm minimizes arclength instead of mean square root residue is due to the existence of background reflection. The arclength minimization is also much more sensitive to ultra-sharp peaks than the mean square root.

The decomposition algorithm can be summarized as follows:

**Input:** Spectrum $R(E)$ with Fano resonances

**Output:** Fano parameters $E_0^{(1)}, E_0^{(2)}, \ldots, E_0^{(n)}, \Gamma^{(1)}, \Gamma^{(2)}, \ldots, \Gamma^{(n)}, \delta^{(1)}, \delta^{(2)}, \ldots, \delta^{(n)}$ and background reflection spectrum $B(E)$.

```
1   begin
2       interpolate the input spectrum $R(E)$ with uniform and sufficient resolution
3       detect and count sharp peaks (count=n) in the input spectrum $R^{(0)}(E)$
4       for i=1, 2, ..., n
5           estimate the initial guess of $E_0^{(i)}, \Gamma^{(i)}\ and, \delta^{(i)}$ based on the $i$th peak location and width
6           minimize eq. (7) with multivariate optimization algorithm to determine $E_0^{(i)}, \Gamma^{(i)}, \delta^{(i)}$
7           if $r\left(E_0^{(i)}, \Gamma^{(i)}, \delta^{(i)}\right) > \sum_{j=1}^{m-1} \sqrt{\left(R^{(i)}(E_{j+1}) - R^{(i)}(E_j)\right)^2 + (E_{j+1} - E_j)^2}$
8               discard current optimization result
9           else
10              $R^{(i)}(E) = R^{(i-1)}(E) - \bar{\sigma}^{(i)}\left(E; E_0^{(i)}, \Gamma^{(i)}, \delta^{(i)}\right)$
11      $B(E) = R^{(n)}(E)$
```

## Data availability

The data that support the scenario within this paper and other findings of this study are available from the corresponding authors upon reasonable request.

## Code availability

All codes including the Adaptive data acquisition (ADA) method, Decomposition algorithm, regression neural networks, and the semi-supervised classification neural networks are available from the corresponding authors upon reasonable request.

## Reference

[1]  Ma, X., et al. Engineering photonic environments for two-dimensional materials. *Nanophotonics* **10,** 1031–1058 (2021).

[2]  Dusanowski, Ł., et al. Purcell-Enhanced and Indistinguishable Single-Photon Generation from Quantum Dots Coupled to On-Chip Integrated Ring Resonators. *Nano Letters* **20,** 6357-6363 (2020).

[3]  Hoang, T. X., et al. Collective Mie Resonances for Directional On-Chip Nanolasers. *Nano Letters* **20,** 5655-5661 (2020).

[4]     Vyatskikh, A., Ng, R. C., Edwards, B., Briggs, R. M. & Greer, J. R. Additive Manufacturing of High-Refractive-Index, Nanoarchitected Titanium Dioxide for 3D Dielectric Photonic Crystals. *Nano Letters* **20,** 3513–3520 (2020).

[5]     Fano, U. Effects of Configuration Interaction on Intensities and Phase Shifts. *Physical Review* **124,** 1866-1878 (1961).

[6]     Limonov, M.F., Rybin, M.V., Poddubny, A.N. & Kivshar, Y.S. Fano resonances in photonics. *Nature Photonics* **11,** 543-554 (2017).

[7]     Argyropoulos, C., Monticone, F., D'Aguanno, G. & Alù, A. Plasmonic nanoparticles and metasurfaces to realize Fano spectra at ultraviolet wavelengths. *Applied Physics Letters* **103,** 143113 (2013).

[8]     Hsu, C.W., Zhen, B., Lee, J., et al. Observation of trapped light within the radiation continuum. *Nature* **499,** 188-191 (2013).

[9]     Lim, W.X., et al. Ultrafast All-Optical Switching of Germanium-Based Flexible Metaphotonic Devices. *Advanced Materials* **30,** 1705331 (2018).

[10]    Yan, C., Yang, KY. & Martin O. Fano-resonance-assisted metasurface for color routing. *Light: Science & Applications* **6,** e17017 (2017).

[11]    Chen, J., He, K., Sun, C., Wang, Y., Li, H. & Gong Q. Tuning Fano resonances with a nano-chamber of air. *Optics letters* **41,** 2145-2148 (2016).

[12]    Lan, S., Rodrigues, S.P., Taghinejad, M. & Cai, W. Dark plasmonic modes in diatomic gratings for plasmoelectronics. *Laser & Photonics Reviews* **11,** 1600312 (2017).

[13]    Gupta, T.D., et al. Self-assembly of nanostructured glass metasurfaces via templated fluid instabilities. *Nature Nanotechnology* **14,** 320-327 (2019).

[14]    Wu, S., et al. Monolayer semiconductor nanocavity lasers with ultralow thresholds. *Nature* **520,** 69-72 (2015).

[15]    Zhu, Y., Hu, X., Huang, Y., Yang, H. & Gong, Q. Fast and Low-Power All-Optical Tunable Fano Resonance in Plasmonic Microstructures. *Advanced Optical Materials* **1,** 61-67 (2013).

[16]    Luk'yanchuk, B., et al. The fano resonance in plasmonic nanostructures and metamaterials. *Nature Materials* **9**, 707-715 (2010).

[17]    Hsu, C.W., Zhen, B., Stone, A.D., Joannopoulos, J.D. & Soljačić, M. Bound States in the Continuum. *Nature Reviews Materials* **1,** 16048 (2016).

[18]    Jiang, J., Chen, M. & Fan, J.A. Deep neural networks for the evaluation and design of photonic devices. *Nature Reviews Materials* https://doi.org/10.1038/s41578-020-00260-1 (2020).

[19]    Ma, W., et al. Deep learning for the design of photonic structures. *Nature Photonics* **15,** 77-90 (2021).

[20]    Fang, J., Swain, A., Unni, R. & Zheng, Y. Decoding Optical Data with Machine Learning. *Laser & Photonics Reviews* **15,** 2000422 (2021).

[21]    Kudyshev, Z. A., Shalaev, V. M. & Boltasseva, A. Machine Learning for Integrated Quantum Photonics. *ACS Photonics* **8,** 34–46 (2021).

[22] Peurifoy, J., et al. Nanophotonic particle simulation and inverse design using artificial neural networks. *Science Advances* **4,** eaar4206 (2018).

[23] Ma, W., Cheng, F. & Liu, Y. Deep-Learning-Enabled On-Demand Design of Chiral Metamaterials. *ACS Nano* **12,** 6326-6334 (2018).

[24] Jiang, J. & Fan, J.A. Global Optimization of Dielectric Metasurfaces Using a Physics-Driven Neural Network. *Nano Letters* **19,** 5366-5372 (2019).

[25] Bayati, E., et al. Inverse Designed Metalenses with Extended Depth of Focus. *ACS Photonics* **7**, 873-878 (2020).

[26] Christensen, T., et al. Predictive and generative machine learning models for photonic crystals. *Nanophotonics* **9,** 4183-4192 (2020).

[27] Xu L, Rahmani M, Ma Y, et al. Enhanced light–matter interactions in dielectric nanostructures via machine-learning approach. *Advanced Photonics* **2**, 026003 (2020).

[28] Liu, Z., Zhu, D., Rodrigues, S. P., Lee, K.T. & Cai, W. Generative Model for the Inverse Design of Metasurfaces. *Nano Letters* **18,** 6570–6576 (2018).

[29] An, S., et al. A Deep Learning Approach for Objective-Driven All-Dielectric Metasurface Design. *ACS Photonics* **6,** 3196–3207 (2019).

[30] Raissi, M., Perdikarism P. & Karniadakis, G.E. Physics-informed neural networks: A deep learning framework for solving forward and inverse problems involving nonlinear partial differential equations. *Journal of Computational Physics* **378,** 686-707 (2019).

[31] Zhu, Y., Zabaras, N., Koutsourelakis, P.S. & Perdikaris, P. Physics-Constrained Deep Learning for High-dimensional Surrogate Modeling and Uncertainty Quantification without Labeled Data. *Journal of Computational Physics* **394,** 56-81 (2019).

[32] Jin, J., et al. Topologically enabled ultrahigh-Q guided resonances robust to out-of-plane scattering. *Nature* **574,** 501-504 (2019).

[33] Zhao, J.Q., Zeng, P., Lei, L.P. & Ma, Y. Initial guess by improved population-based intelligent algorithms for large inter-frame deformation measurement using digital image correlation. *Optics and Lasers in Engineering* **50,** 473-490 (2012).

[34] Lemaréchal, C. Cauchy and the gradient method. *Doc Math Extra* **10,** 251-254 (2012).

[35] Kirkpatrick, S., Gelatt, C.D. & Vecchi, M.P. Optimization by Simulated Annealing. *Science* **220,** 671-680 (1983).

[36] Johnson, S.G. & Joannopoulos, J.D. Block-iterative frequency-domain methods for Maxwell's equations in a planewave basis. *Optics Express* **8,** 173-190 (2001).

[37] Lee, J., et al. Observation and Differentiation of Unique High-Q Optical Resonances Near Zero Wave Vector in Macroscopic Photonic Crystal Slabs. *Physical Review Letters* **109,** 067401 (2012).

[38] Torrey, L. & Shavlik, J. *Handbook of Research on Machine Learning Applications and Trends: Algorithms, Methods, and Techniques Ch.11* (IGI Global, Hershey, 2010).

[39] Ma, X., Zhang, R., Sun, J., Shi, Y. & Zhao, Y. Reduction of Reactive-Ion Etching-Induced Ge Surface Roughness by $SF_6/CF_4$ Cyclic Etching for Ge Fin Fabrication. *Chinese Physics Letters* **32,** 045202 (2015).

**Author contributions**

X. M., Y. M., and S. L. conceived the idea and initiated the project. X.M. and Y.M. performed the numerical simulation. Y.M. built the machine learning models and developed the algorithms. Y.M., X.M., and Q.L developed the analysis codes. X.M. fabricated devices and performed angle-resolved band structure measurements. X.M., Y.M., Q. L., D.X., and J.W. collected the training sets. X.M., Y.M, and S.L. wrote the manuscript with P.C and K.K help. All authors discussed the results and manuscript. S.L. supervised the project.

**Competing interests**

The authors declare no competing interests.

**Additional information**

Supplementary information is available for this paper.

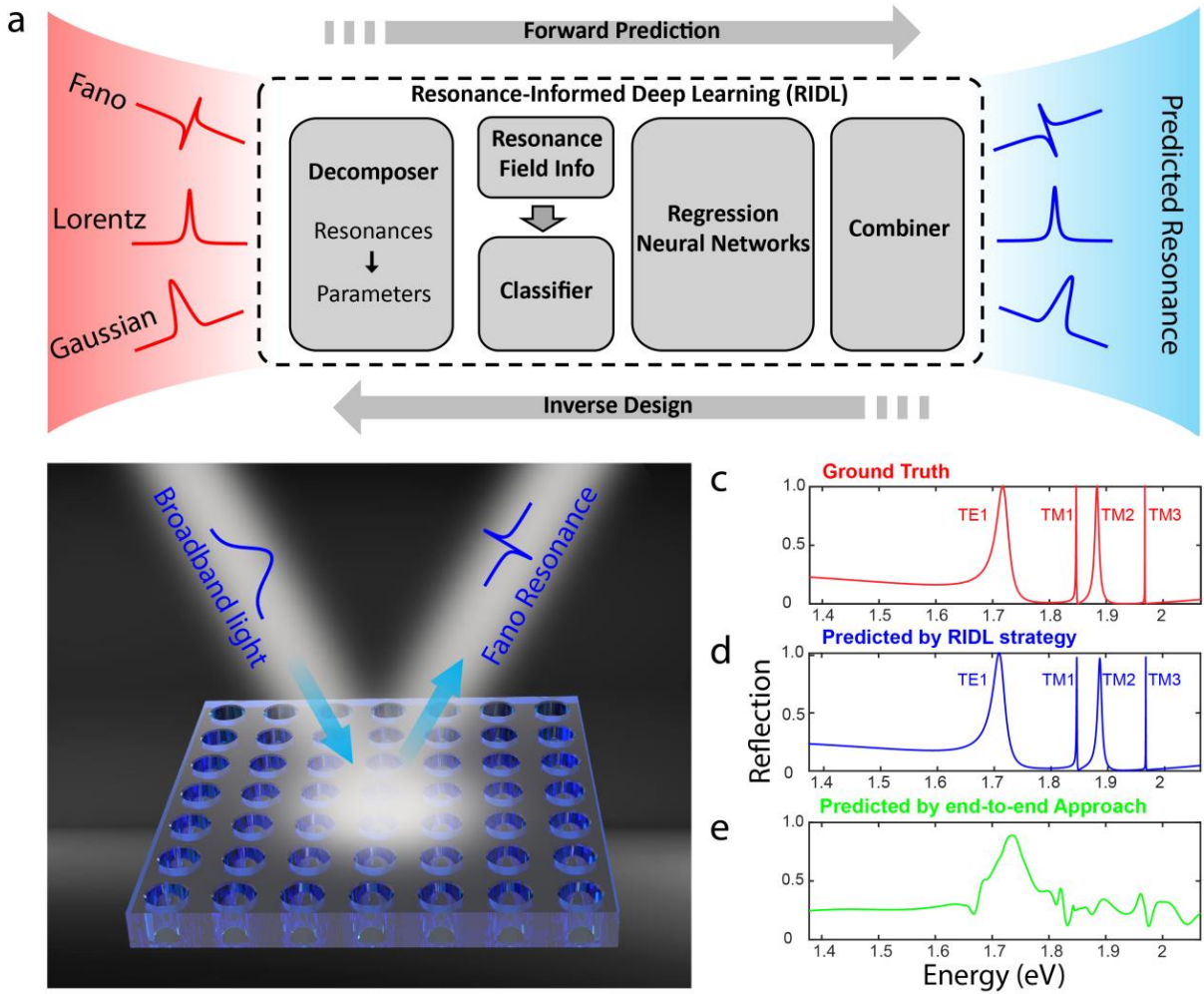

**Figure 1. Resonance-informed deep learning (RIDL) strategy can accurately predict resonances and needs less training datasets. a)** Schematic of the RIDL strategy, illustrating a decomposer that decompose resonances with sharp peak lineshape from their spectra; a classifier that classifies resonances using their field information; regression neural networks that are trained with the classified resonances and a combiner that joins the predicted resonances and background spectra, generating predicted spectra. **b)** As a benchmark platform to demonstrate RIDL, photonic crystal (PhC) slabs made of silicon nitride was selected to generate reflection spectra with Fano resonances. **c-e)** The reflection spectrum of the PhC slab: original spectrum numerically calculated as ground truth, spectrum predicted by the RIDL strategy, and spectrum predicted by a traditional end-to-end deep learning approach, respectively. The spectrum is reflected from a PhC slab with parameters of pitch size a = 500 nm, radius of hole r = 132 nm and thickness of the slab h = 175 nm under the oblique incident light with incident angle β = 1.12°.

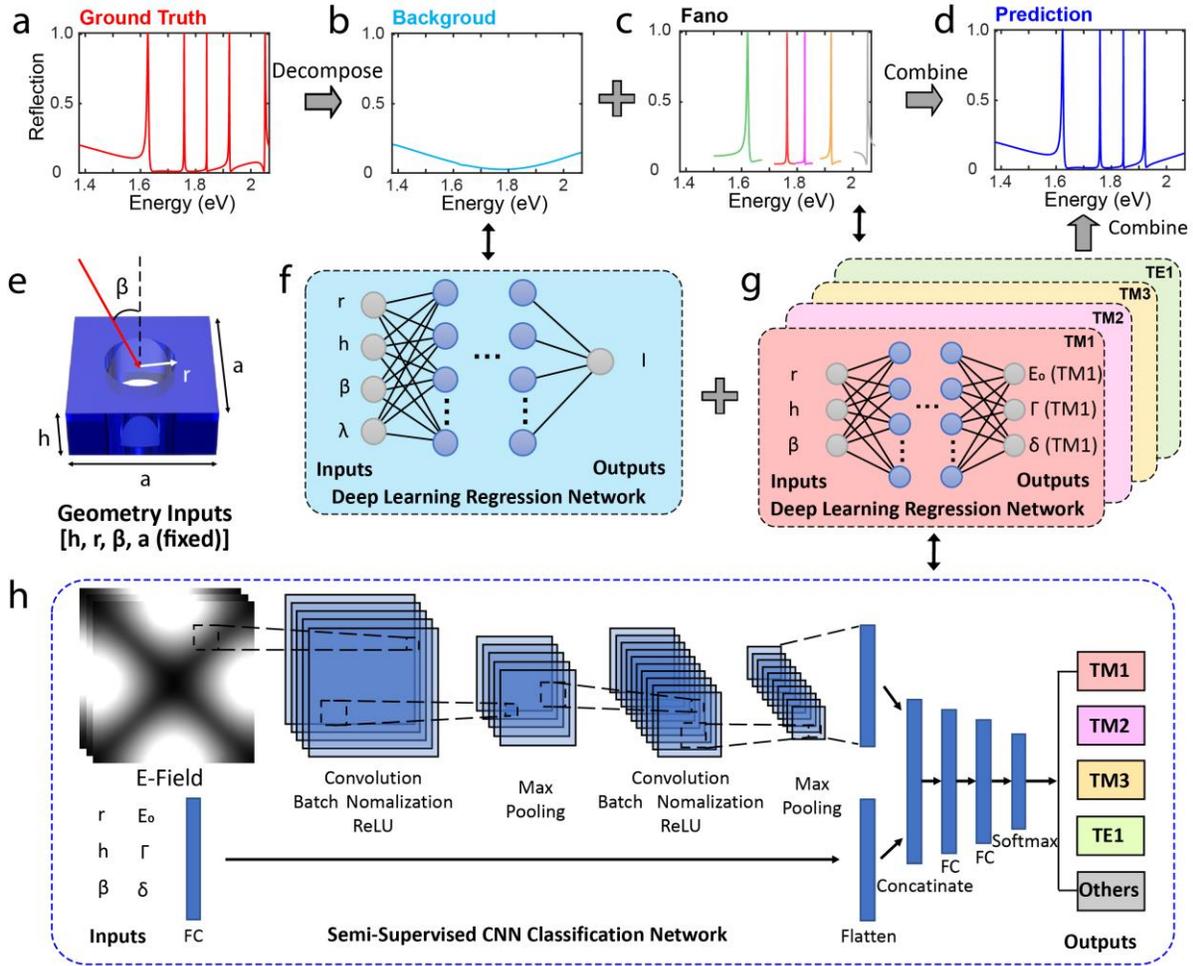

**Figure 2. Details of the RIDL strategy. a-d)** An original reflection spectrum from numerical simulation, the decomposed background and Fano peaks, and the predicted spectrum, respectively. The spectrum is reflected from a PhC slab with parameters of pitch size a=500nm, radius of hole r=89nm and thickness of the slab h=178nm under the oblique incident light with incident angle β=4.9°. **e)** A unit cell of the PhC slab showing the geometry inputs including the pitch size a, radius of holes r, thickness of the slab h and incident angle β. **f)** The deep learning regression network for the background spectra prediction. The inputs include r, h, β, and wavelength in spectra λ and the outputs is the intensity I in spectra for each corresponding λ. **g)** The four parallel sub-deep learning regression networks for four Fano peaks, respectively. In each sub-network, the inputs include r, h, and β and the outputs include a set of Fano parameters include $E_0$, $\Gamma$, and $\delta$. **h)** The CNN classification network, the inputs consists of two branches The first branch is electric field (E-field) strength on the surface of the PhC in the form of three 41 pixel by 41 pixel mapping before peak (E = $E_0$-$\Gamma$/8), on peak (E = $E_0$), and after peak (E = $E_0$ + $\Gamma$/8). The second branch is geometric and Fano parameters of corresponding resonances. The first branch input is treated with convolution and max pooling layers whereas the second branches are treated by full connection (FC) layers. The two branches are merged by a concatenation layer. This classifier separates Fano peaks into five categories, TM1 (the first transverse magnetic mode), TM2, TM3, TE1 (the first transverse electric mode) and others.

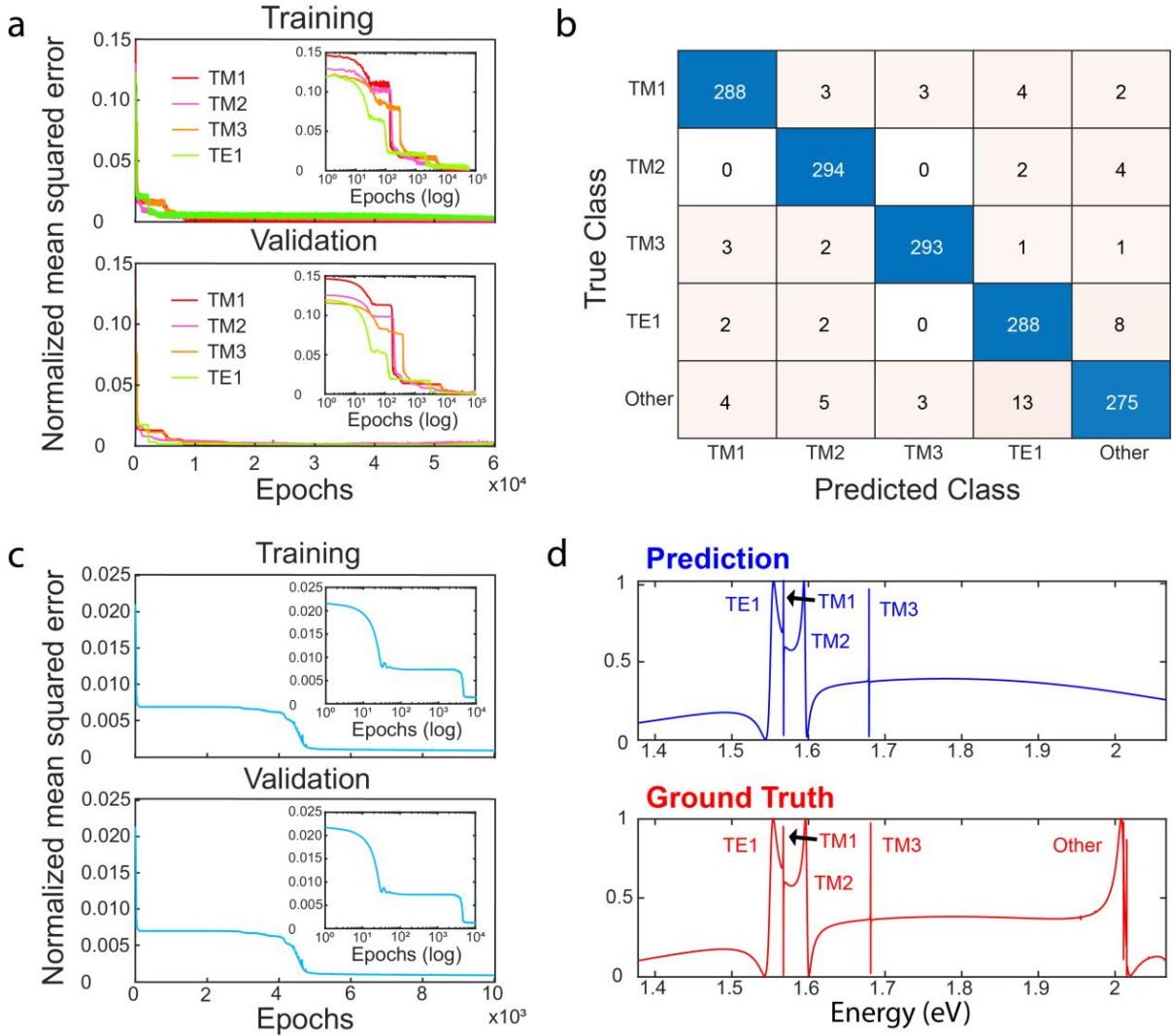

**Figure 3. The performance of the sub-deep learning neural networks.** The normalized mean squared error (MSE) of **a)** the four sub-regression networks for each Fano resonance, and **c)** the MSE of the background regression network. The insets in **a)** and **c)** are the log scale plots to show the convergence trend more clearly. **b)** The confusion matrix of the Fano classification network trained by only 200 labeled Fano peaks from each resonance category. A high accuracy (95%) was achieved on the test set with minimum bias. **d)** The comparison between a prediction spectrum (blue) and the ground truth by numerical simulation (red) to demonstrate the robustness of RIDL. The TM1 resonance with ultra-high Q-factor is situated between two resonance with broad lineshape (TE1 and TM2 resonances). The spectrum is reflected from a PhC slab with parameters of pitch size a = 500 nm, radius of hole r = 108 nm and thickness of the slab h = 275 nm under the oblique incident light with incident angle β = 0.47°.

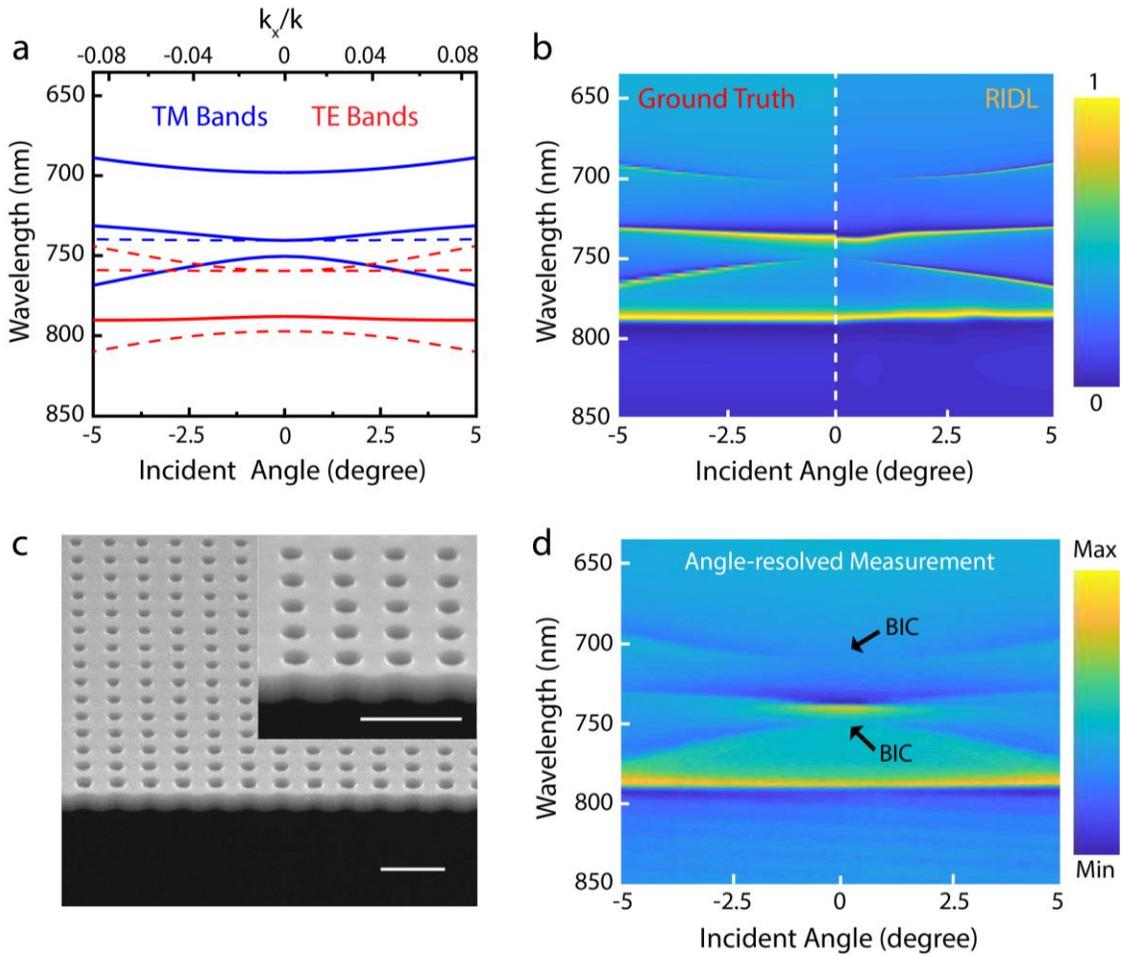

**Figure 4. Inverse design of a PhC slab with two BIC using the RIDL.** The design constraints include **1.** the radius of holes is larger than 80 nm; **2.** the thickness of slab is between 200 nm to 300 nm, and **3.** two BIC modes located at 700 nm and 750 nm, respectively. The designed PhC has pitch size a = 500 nm, radius of hole r = 101 nm and thickness of the slab h = 238 nm. **a)** Band structure using inversely designed parameters by MPB band solver. **b)** The reflection band structure by numerical simulation (left) and RIDL strategy (right). **c)** SEM image of a fabricated PhC slab with the inversely designed parameters. Scale bar: 1 μm. **d)** The experimentally measured angle-resolved reflection band structure of the fabricated PhC slab, the two BIC modes are labeled.